\documentclass[12pt,preprint]{aastex}
\usepackage{emulateapj5,apjfonts}

\shorttitle{It's EZ to Evolve \scshape Zams \upshape Stars}
\shortauthors{Paxton}

\def\Msun{\hbox{$M_{\odot}$}}
\def\Lsun{\hbox{$L_{\odot}$}}
\def\Rsun{\hbox{$R_{\odot}$}}

\begin{document}

\title{It's EZ to Evolve \scshape Zams \upshape Stars:\\ A Program Derived from Eggleton's Stellar Evolution Code}

\author{Bill Paxton}
\affil{Kavli Institute for Theoretical Physics\\
Kohn Hall, University of California, Santa Barbara, CA 93106;
paxton@kitp.ucsb.edu}

\begin{abstract}

``Evolve \scshape Zams\upshape'', ``EZ'' for short, is derived from Peter Eggleton's stellar evolution program.   The core of EZ is a stripped down, rewritten version of a subset of Eggleton's code, specialized to handle single star evolution from the zero-age main sequence until forced to stop by an event such as a helium flash or a crystallizing core.  The procedure and data interfaces to the program are designed to be easy to use while still providing a wide range of function.  EZ is written in Fortran 95 following current programming practices and can be downloaded from  the web site at http://theory.kitp.ucsb.edu/$\sim$paxton/.

\end{abstract}

\section{Introduction}
The main sequence is where stars spend most of their lives while they burn hydrogen in their cores.  The zero-age main sequence (\scshape Zams\upshape) corresponds to stars that are just settling into their careers as hydrogen burners after having formed from a collapsing gas cloud.   After a long time spent doing pretty much the same thing, they eventually evolve away from the main sequence and begin burning other elements.  Stellar evolution programs simulate this based on the relevant opacities and nuclear rates and some simplifying assumptions that hopefully preserve the important features while making the computation tractable.

Plots are a great tool to show stellar structure and its evolution.  The textbooks (e.g., Clayton, 1968, Hansen and Kawaler, 1994, and Kippenhan and Weigert, 1990) and the research journals are full of them -- stars evolving along twisting paths in the Hertzsprung-Russell diagram, convection zones and burning shells moving about at various stages of evolution, or Sun-like stars ascending the giant branch to ignite in a helium flash while similar stars peel off to become white dwarfs.  The plots are fascinating to look at, but it is even better to make them yourself.   What is needed is a simple-to-use evolution package.  

Peter Eggleton's stellar evolution program has much to recommend it as a candidate for this kind of application (descriptions can be found in Eggleton, 1971, Eggleton, 1972, Eggleton et al., 1973, and Eggleton, 1973).  It has been widely used and repeatedly upgraded for a series of projects (recent examples include Han et al., 1994, Pols et al., 1995, and Han et al., 2003).  It has introduced innovations to the field, as evidenced by the discussion of relaxation methods for boundary value problems in Press et al. (1988), which is based on algorithms developed by Eggleton.

``Evolve \scshape Zams\upshape'', ``EZ'' for short, is derived from Eggleton's program as it was in late 2003.   The core of EZ is a stripped down, rewritten version of a subset of Eggleton's code, specialized to handle single star evolution from the zero-age main sequence until forced to stop by an event such as a helium flash or a crystallizing core.  You can download EZ from the web site at http://theory.kitp.ucsb.edu/$\sim$paxton/. The download contains a \scshape readme \upshape file to guide you from there in building and using the system.

The following sections give more details.  Section 2 sketches the changes that were made to get to EZ from Eggleton's program.  Section 3 outlines the new procedure interface, while Section 4 covers the new data interface.  Finally, section 5 describes the demo programs that illustrate what you can get from EZ with a modest amount of work.

\section{Getting to EZ from Eggleton's Code}

Blame it on Lars.  That's Lars Bildsten at the University of California, Santa Barbara.  He suggested looking at the late ignition of helium in stars that were on the way to becoming white dwarfs (along the lines of Han et al., 2003).  It soon became clear that the project needed something better than I could put together from the pieces of code floating around Lars' group.  That led me to e-mail Peter Eggleton, and he kindly sent me his current version.   I spent several months getting to know Eggleton's code and modifying it for use in the project with Lars.   The following is a sketch of that process.

The first step was to replace  \scshape goto\upshape's  by label-less control structures to help me understand the program flow.  This was followed by replacing \scshape implicit\upshape's by explicit type declarations for the variables and procedures.  The program was simplified by removing the parts for binary evolution.  Then it was converted to Fortran 95, and the \scshape common \upshape blocks were eliminated.  Symbolic names replaced numeric constants in the subroutines, and the data declarations were restructured and given extensive comments.  A control interface was created based on a small number of procedures, and a data interface was constructed based on Fortran 95 data \scshape module\upshape's.  Procedures were added to allow a complete state of the system to be saved and restored during a run.  Finally, EZ was extensively tested by comparing low level output with similar data from Eggleton's code. 

\section{Procedure Interface}

EZ can handle stars from 0.1 $\Msun$ up to 100 $\Msun$ with several choices for the initial metallicity from 0.0001 to 0.03.  There is also a routine to let you modify the helium abundance.  The zero-age main sequence stars are created by adjusting the closest of a large set of precomputed models, so getting to an arbitrary starting point is quite fast.   The basic two step operation of EZ is (1) create a \scshape Zams \upshape model of a certain initial mass and composition, and (2) evolve it, passing each model as it is created to a call-back subroutine that you provide which can set system parameters and write to output files.  Along the way, you have the option of saving an intermediate state in order to restore it later.  This can be a great help when you are searching for a parameter value to accomplish some specific outcome and the parameter only influences later stages of evolution (such as rate of envelope ejection).

\section{Data Interface}
The data interface is simply a Fortran module with declarations and lots of comments. The interface includes evolution parameters and their default values.  Most will go untouched, but as needed they can be changed either at initialization or between steps. There are parameters such as $\eta $ for Reimers' wind, with mass loss rate equal to $4 \times 10^{-13} \Msun$/year $( \eta R L / M)/(\Rsun \Lsun / \Msun)$ (Reimers, 1975),  $\alpha $ for convective mixing length, and more.  There are also parameters unrelated to the physics,  such as error tolerances and time step controls.

In addition to legitimate evolution, there are parameters to let you do ``pseudo-evolution''.  You can add or remove mass from the surface at arbitrary rates, you can artificially inject energy, you can turn off convective overshooting, and you can do other arcane things as well.   Such pseudo-evolution can be useful in constructing models.  The third demo uses artificial mass loss to mimic envelope ejection during the common envelope phase of a binary, with the resulting star then allowed to evolve normally after the envelope is gone.

\section{Demo Programs}

The demo programs, which can be downloaded as part of the software distribution from the website, illustrate the use of EZ and provide templates for your applications.  The first demo explores the \scshape Zams \upshape for two different metallicities.  The second demo surveys a wide range of masses evolving from the main sequence for Z=0.02.  The third considers a 1 $\Msun$ star under different scenarios of mass loss including one that leads to a late helium flash when the star is well along the track to becoming a white dwarf.  The web site has details, as well as figures in pdf format for downloading.  Take a look at the on-line figures; they may be useful supplements to the standard textbook fare.  Here are a few more details.

Demo1 computes the zero age main sequences for Z=0.02 and Z=0.0001.  The variation with mass along the \scshape Zams \upshape is shown for luminosity, surface temperature, central temperature, central density, radius, \scshape p-p \upshape versus \scshape cno \upshape nuclear reactions, opacity, nuclear burning time scale, and convection zones.

Demo2 shows the HR diagram tracks and central temperature and density tracks for a sample of Z=0.02 stars over a wide range of initial mass. The tracks are marked to show the location of break-even for the net power from nuclear reactions beyond hydrogen burning minus the total power lost to neutrino cooling (i.e., the onset of significant helium burning).  The demo also looks in detail at a large number of stars, all simulated with convective overshooting and Reimers' wind ($\eta = 1$).  In addition to the usual HR diagram and central temperature-density tracks, the figures include histories of the later values for radius, neutino losses, power from triple-alpha, power from alpha-capture, power from carbon burning, center degeneracy, and total metal fraction.  There are plots that trace the evolution of convection zones, burning zones, central abundances, nuclear power sources, and neutrino cooling.  In addition, there are figures showing sets of profiles by mass coordinate taken at key moments along the evolution.

Demo3 treats a 1 $\Msun$ star with Z=0.02.  It takes the star from the main sequence and saves state near the tip of the giant branch.  Then it continues the evolution in three different ways.  First, it evolves with no mass loss up to the helium flash.  Then, it restores the saved state and evolves the star again with the envelope ejected quickly.  This gives a smaller core and produces a white dwarf.  Finally, it restores state and tries again with a reduced rate of mass loss, which gives a slightly larger core and a late helium flash.

\section{Conclusion}

If you would like to learn more about EZ, the next step is to visit the website.  Even if you don't plan to download the code, you might enjoy looking at some of the figures.  If you do decide to download it, there is a tar file with data and source code and a \scshape readme \upshape file to help you along.  It includes instructions for building and running EZ, for testing the results, and for making your own applications.   Please e-mail me with comments, questions, and suggestions for improvements.

Lars Bildsten, at the University of California, Santa Barbara, Kavli Institute for Theoretical Physics (KITP), has been a teacher, advisor, and friend for a retired computer scientist with a desire to learn some astrophysics.  Phil Arras, a post-doc  at KITP, kindly provided code for neutrino cooling (based on Itoh et al., 1996) and volunteered to be the first beta-user.  And special thanks to Peter Eggleton himself for providing the current version of his code and e-mail support while I was getting into it.  This work was partially supported by the National Science Foundation under grants PHY99-07949 and AST02-05956.

\end{document}